\documentclass[superscriptaddress,twocolumn,
amssymb,amsmath,nobibnotes,aps,pre,showkeys,showpacs]{revtex4-1}
\usepackage{bm}
\usepackage{bbm}
\usepackage{latexsym}
\usepackage{dcolumn}
\usepackage{amsfonts,amssymb,amsmath}
\usepackage{graphicx,epsfig}
\usepackage{subfigure}
\usepackage{color}
\usepackage{ulem}
\usepackage{amsmath}

\begin{document}
	
\title{Survival of current in a periodically driven hard-core bosonic system}
\author{Rashmi J. Sharma}\email{jangid.rashmi@gmail.com}\affiliation{Department of Physics, Birla Institute of Technology and Science, Pilani 333031, India} 
\author{Jayendra N. Bandyopadhyay}\email{jnbandyo@gmail.com}\affiliation{Department of Physics, Birla Institute of Technology and Science, Pilani 333031, India} 

\begin{abstract}
We study the survival of the current induced initially by applying a twist at the boundary of a chain of hard-core bosons (HCBs), subject to a periodic double $\delta$-function kicks in the staggered on-site potential. We study the current flow and the work-done on the system at the long-time limit as a function of the driving frequency. Like a recent observation in the HCB chain with single $\delta$-function kick in the staggered on-site potential, here we also observe many dips in the current flow and concurrently many peaks in the work-done on the system at some specific values of the driving frequency. However, unlike the single kicked case, here we do not observe a complete disappearance of the current in the limit of a high driving frequency, which shows the absence of any dynamical localization in the double $\delta$-functions kicked HCB chain. Our analytical estimations of the saturated current and the saturated work-done, defined at the limit of a large time together with a high driving frequency, match very well with the exact numerics. In the case of the very small initial current, induced by a very small twist $\nu$, we observe that the saturated current is proportional to $\nu$. Finally, we study the time-evolution of the half-filled HCB chain where the particles are localized in the central part of the chain. We observe that the particles spread linearly in a light-cone like region at the rate determined by the maximum value of the group velocity. Except for a very trivial case, the maximum group velocity never vanishes, and therefore we do not observe any dynamical localization in the system. 
\end{abstract}

\maketitle
	
\section{Introduction}

Periodically driven classical and quantum systems are ubiquitous in different areas of physics which include atomic and molecular physics \cite{moore,cold_atom}, quantum optics \cite{qoptics}, nonlinear dynamics \cite{kicked-rotors, kapitza, moore}, electrodynamics \cite{emt-floquet}, etc. For example, time-periodic Hamiltonian is needed to describe the behavior of any material under the illumination of a laser source or to explain the working principle of any physical device connected to an AC source \cite{Oka}. Recently, besides the uses of periodic driving to excite, to control, and to probe different physical systems, periodically driven systems have got its prominence in the condensed matter community because of its possibilities in engineering new phases of matter having exotic quantum properties. For example, graphene under the presence of periodic driving has revealed rich topological textures \cite{Oka, Takashi, Kitagawa, Gu, Morell, Campbell, Delplace, Usaj, Foa, P.M, Sentef}. Some of these exotic quantum phases of matters can only be engineered by periodic driving because they do not have any equilibrium counterpart. Recent developments in the field of ultra-cold atoms, particularly in designing optical lattices of desired potentials, have provided a flexible platform for the experimental realization of these exotic quantum matters \cite{cold_atom}. See Ref. \cite{bukov} for an extensive review on different classes of the periodically driven systems. 

These periodically driven systems are theoretically studied under Floquet formalism; hence, these systems are also known as the Floquet systems. Also, the process by which a periodic driving controls a quantum system and produces exotic phases of matters is called {\it Floquet engineering}. The most important element of Floquet engineering is the Floquet Hamiltonian $H_{t_0}^F$. This is a time-independent effective Hamiltonian, which can reproduce the time-evolution $U(t_0, t_0+T)$ generated by a driven Hamiltonian satisfying the time-periodic property $H(t+T) = H(t)$ over one driving cycle $T$, where $U(t_0, t_0+T) = \exp\left(- \frac{i}{\hbar} T H_{t_0}^F \right)$. At the fundamental level, a major challenge of Floquet engineering is to design a driving scheme to construct the time-period Hamiltonian $H(t)$ such that one can get a desired effective static Hamiltonian which is also known as the Floquet Hamiltonian. On the other hand, at the theoretical side, a major challenge is to calculate the Floquet Hamiltonian for a given time-periodic Hamiltonian. Except for a very few cases, one has to employ a perturbation scheme to obtain the Floquet Hamiltonian approximately. 

In this paper, we are considering one such exceptional cases where we can derive the Floquet Hamiltonian exactly. Here we study the dynamics of a chain of hard-core bosons (HCBs) which is driven periodically in time through the staggered onsite potential. In the case of HCBs, strong repulsive interactions between the bosons prevent them from occupying the same lattice site, and thus they mimic the Pauli exclusion principle for fermions. However, when two bosons are placed at two different sites, then they behave like normal bosons and satisfy the standard bosonic algebra with characteristic momentum distribution. Recently, the HCBs have also been observed experimentally in optical lattice \cite{Paredes, Kinoshita}. This experimental observation led to an extensive study of the bosonic system in one-dimension \cite{Cazalilla}. Recently, the dynamics of HCBs is studied under the sinusoidally and $\delta$-function kicked driven staggered onsite potential \cite{Tanay, sinusoidal}. Here, the HCB system is driven periodically in the form of double $\delta$-function kicking. Because of the $\delta$-function kicking, one can factorize the Floquet time-evolution operator as a product of simpler unitary operators (see for example Ref. \cite{qchaos}). For example, in the case of a single kicked HCB system, the Floquet time-evolution operator can be factorized into a product of two unitary operators, where one of them describes the free hopping and the other one describes the staggered onsite potential. Mathematically this is equivalent to the $2$-steps modulation scheme which is a special case of the recently proposed $N$-step modulation scheme for the Floquet engineering \cite{dalibard}. The HCB system with periodically kicked staggered potential has recently been studied in the context of dynamical localization (DL) \cite{Tanay}. Here we are considering a {\it double} $\delta$-functions kicked version of the HCB (DKHCB) system. In this case, the staggered onsite potential is introduced to the system in the form of a pair of $\delta$-function kicks within one time-period $T$ and the kicks are separated by some time interval $(< T)$. Here the magnitude of the driving strength of both the kicks is chosen equal but of the opposite sign. This double kicked protocol is equivalent to the $5$-steps modulation scheme proposed in Ref. \cite{dalibard}. We are studying the DKHCB system due to the following reasons: (1) Recent studies have shown that, in many situations, the behavior of the double kicked systems can be drastically different from its single kicked counterpart \cite{Jiao,JNB,fractal}; (2) The double kicked scheme is important because it has already been identified as one of the schemes to create a synthetic magnetic field by shaking specially designed quantum systems, such as cold atomic gases trapped in optical lattices \cite{Creffield}.                

This paper is organized in the following way. Sec. \ref{formalism} is divided into two major parts. First, we introduce our model of the periodically driven hard-core bosons. Then we have formalized the problem from the Floquet theory point of view. In Sec. \ref{results}, we have presented exact numerical results which are supported very well by our  analytical estimation. Finally, we conclude in Sec. \ref{conclude}.
 
\section{The model and the Floquet formalism}
\label{formalism}

Hamiltonian for a chain of hard-core bosons (HCBs) on a lattice at half filling is written as:
\begin{equation}
H= - w \sum_{l=1}^L (b_{l}^{\dagger}b_{l+1} + b_{l+1}^{\dagger}b_{l})
\end{equation}
where $b_{l}^{\dagger}$ and $b_{l}$'s are the bosonic creation and annihilation operators, respectively. The creation operator $b_{l}^{\dagger}$ increases the number of bosons at the site $l$ by {\it one}, whereas the annihilation operator $b_l$ does the opposite by lowering the number of bosons {\it one} at the site $l$. These operators satisfy the bosonic commutation relation $[b_l, b_{l'}^{\dagger}] = 0$ and the hard-core relations $(b_l)^2 = (b_l^\dagger)^2 = 0$. The hard-core relations ensure the presence of at the most one boson per lattice site. The parameter $w$ determines the strength of the hopping between two neighboring sites. The HCBs are ordinary bosons, but repel strongly like fermions when they come to the same lattice site. Because of this particular similarity, the HCBs can be mapped in to the non-interacting spin-less fermions with the help of Jordan-Wigner transformations \cite{JW}:
\begin{equation}
\begin{split}
b_{l}= \exp\biggl[-i\pi\sum_{l^{'}=1}^{l-1}f_{l^{'}}^{\dagger}f_{l^{'}}\biggr]f_{l},\\
b_{l}^\dagger=f_{l}^\dagger \exp\biggl[i\pi\sum_{l^{'}=1}^{l-1}f_{l^{'}}^\dagger f_{l^{'}}\biggr],
\end{split}
\end{equation} 
where $f_l^\dagger$ and $f_l$s are fermionic creation and annihilation operators. By the Fourier transformation, this Hamiltonian can be represented in the momentum space. In this space, the Hamiltonian gets decoupled into $2 \times 2$ matrix in terms of the momenta $k$ and $k+\pi$, where $-\pi/2\le k \le \pi/2$. We assign the basis vectors $|k\rangle = (1\ 0)^{T}$ and $|k+\pi\rangle = (0\ 1)^{T}$, then the $2 \times 2$ Hamiltonian can be written as (we now set $w=1$ for the rest of the paper):
\begin{equation}
H_k = - 2 \sigma_{z} \cos\,k, 
\end{equation} 
where $\sigma_z$ denotes the $z$-component of the Pauli matrices. For the half filling case, all the momentum states, i.e. $-\pi/2 \le k \le \pi/2$, are filled. Therefore, the ground state for every $k$ mode is the state $(1\ 0)^{T}$ which is the {\it up} eigenstate of the pseudo-spin operator $\sigma_z$.

If the above Hamiltonian is perturbed by a staggered on-site potential of the form 
\begin{equation}
V = - \alpha \sum_{l=1}^L (-1)^{l}\, b_l^{\dagger} b_l,
\label{eq:staggered}
\end{equation}
then this generates a coupling between the modes with momenta $|k\rangle$ and $|k+\pi\rangle$. In this momentum space, the staggered potential becomes
\begin{equation}
V_k = -\alpha \sigma_x.
\end{equation}
The above form shows an opening up of gap in the momentum space at $k=\pm\pi/2$. This implies that the introduction of the staggered potential is responsible for a quantum phase transition from the gapless superfluid phase at $\alpha = 0$ to the gapped Mott insulator phase for any $\alpha \ne 0$.   

Here we are interested in studying the fate of an initial current in the superfluid phase under the influence of the staggered potential. However, the ground state of the above Hamiltonian has zero current. Therefore, we apply a boost of $\nu$ to the system, and the corresponding Hamiltonian becomes:
\begin{equation}
H^{(\nu)} = - \sum_{l=1}^L\biggl(b_{l}^{\dagger}b_{l+1}e^{-i\nu}+b_{l+1}^{\dagger}b_{l}e^{i\nu}\biggr).
\end{equation}
The ground state of this boosted Hamiltonian has a nonzero current. Instead of keeping the $\nu$-dependent phase factor at each term, one can remove those in the boosted Hamiltonian by the transformation: $b_l \rightarrow e^{i \nu l} b_l$. Here we are assuming periodic boundary condition, which implies that the last hopping term in the (boosted) Hamiltonian represents a hop from the last site $L$ to the site $1$. Therefore, the above transformation does not remove the phase factor in the last term, and the acquired phase factor is $e^{i \nu L}$. Therefore, this is equivalent to a {\it twist} given at the boundary. In the momentum space, for each $k$-mode, the boosted/twisted Hamiltonian becomes
\begin{equation}
H_k^{(\nu)} = - 2 \sigma_z \cos(k-\nu).
\label{eq:ham_twist}
\end{equation}
This indicates that the boost effectively shifts the momentum from $k$ to $k-\nu$. This shift in the momenta or $k$-modes results in the shift in the ground state. Now for the boosted (or twisted) Hamiltonian, the ground state is in the range $-\pi/2 \le k \le -\pi/2+\nu$ and it becomes the state $(0,\,1)^T$; whereas for the range $-\pi/2+\nu \le k \le \pi/2$, the ground state remains to be the state $(1,\, 0)^T$. This asymmetry in the $k$-space leads to a nonzero current in the ground state of the boosted Hamiltonian. 

The observable corresponding to the current is defined by the following Hermitian operator:
\begin{equation}
\widehat{j} = - \frac{1}{L} \left(\frac{\partial H_{\nu}}{\partial \nu}\right)_{\nu=0} =  \frac{i}{L} \sum_{l=1}^L \left(b_{l+1}^{\dagger} b_{l} - b_{l}^{\dagger} b_{l+1}\right).
\end{equation}
In the $k$-space, for each $k$-mode, the current operator will be
\begin{equation}
\widehat{j}_k = \frac{2}{L}\, \sigma_z \sin k.
\label{eq:curr_defn1}
\end{equation} 
In the thermodynamic limit ($L \rightarrow \infty$), the ground state of the boosted system carries a non-zero current of the amount 
\begin{equation}
J_{\mbox{in}} = \frac{2}{\pi}\, \sin\nu.
\label{eq:Jin}
\end{equation}
We prepare the initial state of the system in the current carrying ground state of the boosted/twisted Hamiltonian. We then remove the twist in the Hamiltonian by setting $\nu = 0$ and evolve the initial current carrying state by the un-twisted Hamiltonian $H_k$ perturbed by a periodically driven staggered onsite potential as given in Eq. \eqref{eq:staggered}. 

Here we consider the periodic driving in the form of a pair of Dirac $\delta$-functions within the time-period $T$ as given below:
\begin{widetext}

\begin{figure}[t]
\begin{tabular}{lc|r}
\includegraphics[height=5cm,width=7cm]{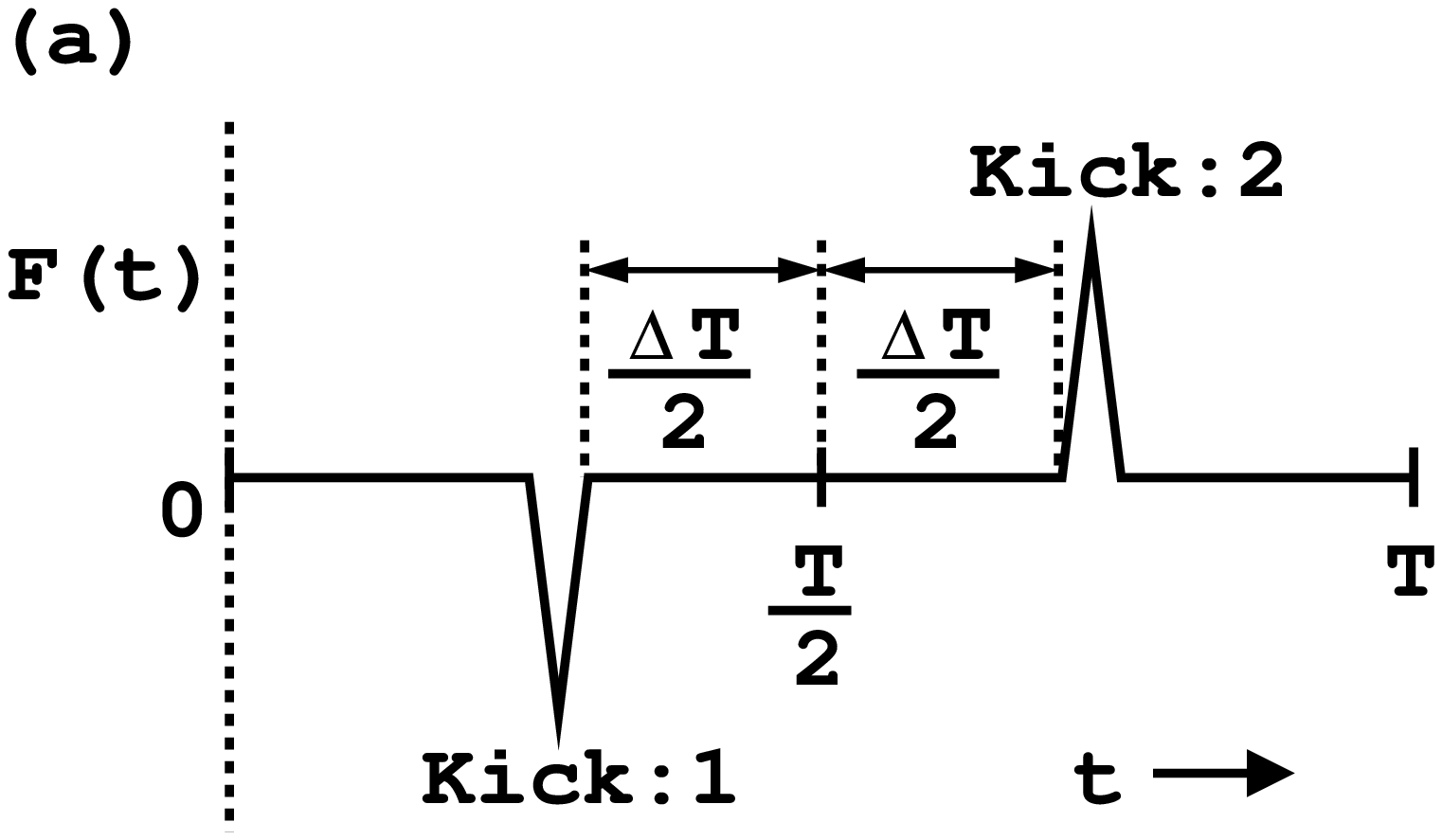}
&&
\includegraphics[height=5cm,width=10cm]{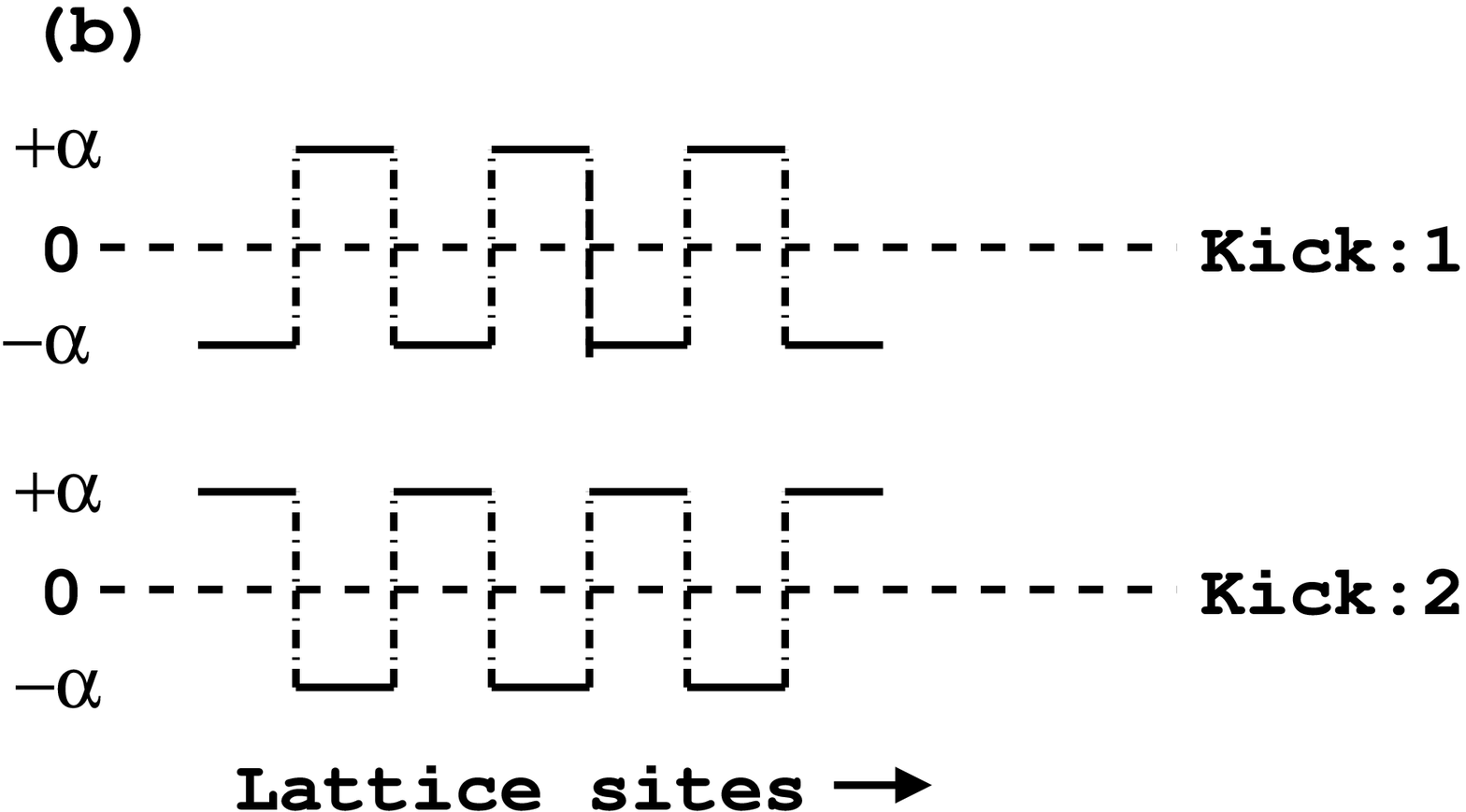}
\end{tabular}
\caption{(a) The driving scheme is shown within the time-period $[0, \,T]$. This driving scheme will repeat itself in between any arbitrary interval of $nT$ and $(n+1)T$. (b) The shape of the staggered potentials are shown at the time of two kicks (at $t_{\mp} = \frac{1}{2}(1 \mp \Delta) T$).}
\label{fig:fig0}
\end{figure}

\begin{equation}
\begin{split}
&V(t) = - \alpha\, F(t) \sum_{l=1}^L (-1)^{l}\, b_l^{\dagger} b_l,\\
{\rm where}\,\,\, F(t) &= \sum_{n=1}^\infty \left\{\delta\left[t - \left(2n-1-\Delta\right)\frac{T}{2}\right] - \delta\left[t - \left(2n-1+\Delta\right)\frac{T}{2}\right] \right\}.
\end{split}
\label{eq:driving}
\end{equation}  
\end{widetext}
Fig. \ref{fig:fig0}(a) shows that the pair of kicks are acting symmetrically on the system about the time $\frac{T}{2}$ at $t = T_{\mp} = (1 \mp \Delta) \frac{T}{2}$. The sign of the two kicks are opposite to each other and because of that the staggered potentials are negative of each other. These potentials will overlap with each other if we shift one of the potentials by a single lattice unit.

In the momentum space $(k,\,k+\pi)$, the DKHCB system can be written as: 
\begin{equation}
H_k(t) = - 2 \sigma_{z} \cos k - \alpha\, \sigma_x F(t), 
\label{eq:p_sp-Ham}
\end{equation}
where the time-dependent function $F(t)$ is given in Eq. \eqref{eq:driving}. The Floquet time-evolution operator corresponding to the above driven Hamiltonian is constructed from the definition $\mathcal{F}_k(T) = \mathcal{T} e^{-i\int_0^T H_k(t) dt}$, where $\mathcal{T}$ denotes the time-ordered operator, as 
\begin{equation}
\begin{split}
\mathcal{F}_k(T) &= \exp\bigl[i\, T (1-\Delta) \sigma_z \cos k \bigr]\, \exp\left(i\, \alpha \sigma_x\right)\\ &\times \exp\left(i\, 2\, T \Delta  \sigma_z \cos k\right)\, \exp\left(- i\, \alpha \sigma_x \right)\\ &\times \exp\bigl[ i\, T (1-\Delta) \sigma_z \cos k\bigr].
\end{split}
\label{eq:FkT}
\end{equation}
We define a time-independent effective Hamiltonian $H_{\mbox{eff}}$ such that $\mathcal{F}_k(T) = \exp(-i H_{\mbox{eff}} T)$. Here, using a well known identity $e^{ia(\vec{\sigma}.\widehat{n})}e^{ib(\vec{\sigma}.\widehat{m})}=e^{iq(\vec{\sigma}.\widehat{l})}$ where $ \cos\,q= \cos\,a\,\cos\,b-\widehat{n}.\widehat{m}\sin\,a\,\sin b$ and $\widehat{l}=(1/\sin q )\bigl(\widehat{n}\, \sin a\, \cos b +\widehat{m}\,\sin b\, \cos a - \widehat{n}\times\widehat{m}\, \sin a\, \sin b\bigr)$, we can write down the effective Hamiltonian exactly as $H_{\mbox{eff}} = \mu_k \left(\overrightarrow{\sigma}.\,\widehat{l}\right)$ where
\begin{equation}
\begin{split}
\mu_k &= \pm \frac{1}{T} \cos^{-1}\left[\cos^2\alpha\,\cos \left(2T \cos k\right)\right.\\  &+ \left. \sin^2\alpha \cos \left\{2T(1-2\Delta)\cos k\right\}\right] \equiv \mu_k^{\pm},
\end{split}
\label{eq:mu_k}
\end{equation} 
the quasi-energies $\mu_k$ are also the eigenvalues of $H_{\mbox{eff}}$ and the components of the unit vector $\widehat{l}$ are obtained as:
%\begin{widetext}
\begin{equation}
\begin{split}
l_x&=0,\\
l_y&=\frac{1}{\sin (\mu_k T)} \sin (2\alpha) \sin\left(2T\Delta\cos k\right),\\
l_z&=\frac{1}{\sin (\mu_k T)}\biggl[\cos^2 \alpha\,\sin (2T\cos k)\biggr.\\ &+ \biggl.\sin^2 \alpha\,\sin\left\{2T(1-2\Delta)\cos k\right\} \biggr].
\end{split}
\label{eq:l_compo}
\end{equation}

The unitary operator $\mathcal{F}_k(T)$ can be expressed exactly in terms of a $2 \times 2$ matrix of the form
\begin{equation}
\mathcal{F}_k(T) = \begin{bmatrix}
a_k & b_k\\-b_k & a_k^*
\end{bmatrix}
\label{eq:Floquet_Fk}
\end{equation}
where
\[\begin{split} a_k &= e^{i\, 2T \cos k} \left[\cos^2 \alpha + \sin^2 \alpha\, e^{-i 4 T \Delta \cos k}\right] \\ b_k &= \sin(2\alpha) \sin\left(2 T \Delta \cos k\right).\end{split}\]
The above expression suggests that for some specific values the driving strength $\alpha=\pm m\pi, \,m \in \mathbb{Z}$, the $\delta$-function kicks do not affect the temporal evolution of the HCB chain. As a consequence, the initial state remains unchanged under time-evolution and that leads to a constant current flow in the system throughout the time-evolution.
At $\alpha=0$ and $\alpha=\pi$, the Floquet operators are identical; therefore, all the physical properties are expected to be the same at these values of the driving strength $\alpha$. Consequently, the Floquet operator $\mathcal{F}_k(T)$ itself and properties of all the observables of the system will repeat themselves within every interval $m\pi \leq \alpha \leq (m+1)\pi,\,m \in \mathbb{Z}$. Interestingly, at $\alpha = \pi/2$, the Floquet operator becomes the Hermitian conjugate of the Floquet operator corresponding to $\alpha = 0$ or $\pi$. Furthermore, Eq. \eqref{eq:Floquet_Fk} suggests that if we make a transformation $\alpha \rightarrow \alpha + \frac{\pi}{2}$ where $\alpha \leq \pi/2$, then the Floquet operator transforms into its Hermitian conjugate, i.e., $\mathcal{F}_k(T) \rightarrow \mathcal{F}_k^\dagger(T)$. This suggests that if we plot any local observable as a function of $\alpha$, then that function should show a {\it mirror} symmetry about $\alpha = \pi/2$. Later we shall show that this is indeed the case, and the local observables like the current flow and the work-done on the system follow this symmetry property.
 
We now evolve the initial current carrying state under the Floquet time-evolution operator $\mathcal{F}_k(T)$. Except at the time of two $\delta$-function kicks, the HCB chain remains in the superfluid phase. At the time of each kick, the staggered potential is applied and that creates a gap in the energy band which indicates a phase transition in the system and the system becomes a Mott insulator. Therefore, during the free motion, the current flow remains fixed without any drop; and at the time of each kick, the system momentarily becomes a Mott insulator which leads to a drop in the current flow. We like to study the effect of the periodically kicked staggered onsite potential on the initial current flow. For the single kicked case, a recent study has reported a complete disappearance of the current flow in the system at the asymptotic limit of time and also at the limit of large frequency \cite{Tanay}. Here we are considering a driving protocol which is of the form of double $\delta$-functions kick. The current flow through the system after the time $t=nT$ is given as
\begin{equation}
\begin{split}
J(nT) = & \sum_{k} J_{k}(nT)\\ = & \sum_k \langle\Psi_{k}(nT)|\widehat{j}_{k}|\Psi_{k}(nT)\rangle\\ =& \sum_k \langle\Psi_k(0)|\bigl(\mathcal{F}_k^{\dagger}\bigr)^n\,\widehat{j}_k\,\bigl(\mathcal{F}_k\bigr)^n|\Psi_k(0)\rangle,
\label{eq:curr_defn2}
\end{split}
\end{equation}
where $\widehat{j}_k$ is defined in Eq. \eqref{eq:curr_defn1} and $|\Psi_{k}(0)\rangle$ is the ground state of the Hamiltonian $H_k^{(\nu)}$.

We have studied another local observable, that is the work-done $W_d$ on the system. This is a  measure of the amount of energy absorbed by the system from the external driving. The work-done is defined, after the $n$ time-period, i.e., at $t = nT$ as:
\begin{equation}
W_d(nT) = \frac{1}{L}\sum_k W_k(nT) \equiv\frac{1}{L}\sum_{k}\bigl[e_k(nT)-e_k(0)\bigr]
\end{equation}
where $e_k(nT)=\langle\Psi_k(nT)|H_k|\Psi_k(nT)\rangle$ and $e_k(0)=\langle\Psi_k(0)|H_k|\Psi_k(0)\rangle$. Therefore, we can also write down the above expression as:
\begin{widetext} 
\begin{equation}
\begin{split}
W_d(nT) &= -\frac{2}{L}\sum_k\bigl[\langle\Psi_k(nT)|\sigma_z|\Psi_k(nT)\rangle-\langle\Psi_k(0)|\sigma_z|\Psi_k(0)\rangle\bigr]\\
&= -\frac{2}{L} \sum_k \bigl[\langle\Psi_k(0)|\bigl(\mathcal{F}_k^{\dagger}\bigr)^n\,\sigma_z\,\bigl(\mathcal{F}_k\bigr)^n|\Psi_k(0)\rangle - \langle\Psi_k(0)|\sigma_z|\Psi_k(0)\rangle\bigr]
\label{eq:wd_defn2}
\end{split}
\end{equation}
%\end{widetext}

\section{Results}
\label{results}

We now present our results obtained by the exact numerical method and also confirmed those by analytical means. Notably, we are interested in studying the behavior of two local observables: the current flow through the system and the work-done on the system.

%\begin{widetext} 
\subsection{Current flow}

In order to calculate the current flow through the system, first we have to calculate $\bigl[\mathcal{F}_k(T)^{\dagger}\bigr]^n\,\widehat{j}_k\,\bigl[\mathcal{F}_k(T)\bigr]^n$. Using the relation $\mathcal{F}_k (T) = \exp[-i\mu_k (\vec{\sigma}.\widehat{l})T]$ and the expression of $\widehat{j}_k$ given in Eq. \eqref{eq:curr_defn1}, we obtain the expression of the current flow as
\begin{equation}
\begin{split}
J(nT)&= \frac{2}{L}\sum_{k} \biggl[1-2(1-l_z^2)\sin^2(\mu_k nT)\biggr]\,\langle\Psi_{k}(0)|\sigma_{z}|\Psi_{k}(0)\rangle \,\sin\,k\\
&=\frac{2}{L}\sum_{k} f(k)\, \langle\Psi_{k}(0)|\sigma_{z}|\Psi_{k}(0)\rangle \,\sin\,k
\end{split}
\end{equation}
where 
\begin{equation}
f(k)=[1-2(1-l_z^2)\sin^2(\mu_k nT)].
\label{eq:fk}
\end{equation}
\end{widetext}
We are interested in the behavior of the current at the thermodynamic limit $L \rightarrow \infty$. At this limit, we can replace the summation $\frac{1}{L} \sum_k$ by the integration $\frac{1}{2\pi} \int dk$. Thus we get,
\begin{equation}
J(nT) = \frac{1}{\pi} \int\limits_{-\pi/2}^{\pi/2} dk\, f(k)\, \langle\Psi_k(0)|\sigma_z|\Psi_k(0)\rangle \,\sin\,k.
\label{eq:JnT-cont}
\end{equation}

At the large frequency limit $\omega \gg 1$ (or $T \ll 1$), the components of the unit vector $\widehat{l}$ become
\begin{equation}
\begin{split}
l_x &= 0,\\
l_y &=\frac{\Delta \sin(2\alpha)}{\sqrt{\cos^2\alpha + (1-2\Delta)^2\sin^2\alpha}},\\
l_z &=\frac{\cos^2 \alpha + (1-2\Delta)\sin^2\alpha}{\sqrt{\cos^2\alpha + (1-2\Delta)^2\sin^2\alpha}}.
\label{eq:lz}
\end{split}
\end{equation}
The above expression is valid for all the possible values of $\alpha$ and $\Delta$, except for a special case when these parameters concurrently get the values $\alpha = \pi/2$ and $\Delta = 0.5$. One can see that, for this special case, the above expression of $l_z$ becomes indeterminate ($0/0$ form). Here we note that, for $\alpha = \pi/2$, the Floquet time-evolution operator becomes $e^{i 2T (1-2\Delta)\sigma_z \cos k} = e^{-i (1-2\Delta) H_k T}$, and therefore the effective Hamiltonian is 
\[H_{\mbox{eff}} = (1-2\Delta)\, H_k = - 2 (1-2\Delta)\,\sigma_z \cos k.\] 
This implies that the quasi-energies $\mu_k = -2(1 - 2 \Delta) \cos k$ and the unit vector $l_z = 1$. Now, when $\Delta = 0.5$, $H_{\mbox{eff}} = 0$ is just the {\it null} operator and the time-evolution operator will be $\mathcal{F}_k(T) = \mathbbm{1}$ where $\mathbbm{1}$ is the identity operator. This suggests an absence of any dynamics with time-evolution, and therefore the initial state remains as it is. As we predicted earlier, the current flow in the system remains constant at its initial value which can be calculated mathematically using Eqs. \eqref{eq:fk} and \eqref{eq:JnT-cont} and substituting there $l_z = 1.0$ or $f(k) = 1.0$, and thus we get
\begin{equation}
\begin{split}
J(nT) &= \frac{1}{\pi} \int\limits_{-\pi/2}^{\pi/2} dk\, \langle\Psi_k(0)|\sigma_z|\Psi_k(0)\rangle \,\sin\,k\\  &= \frac{1}{\pi} \left[-\int\limits_{-\frac{\pi}{2}}^{-\frac{\pi}{2} + \nu} dk \sin k + \int\limits_{-\frac{\pi}{2}+\nu}^{\frac{\pi}{2}} dk \sin k \right]\\&= \frac{2}{\pi}\, \sin \nu \\ &= J_{\mbox{in}}.
\end{split}
\end{equation}    

At the asymptotic limit $n \rightarrow \infty$, such that $n T \gg 1$, we can replace $\sin^2(\mu_k nT)$ by its average value over one single cycle, that is $\overline{\sin^2(\mu_k nT)} = 0.5$. Thus we get $f(k) = l_z^2$. At this limit, we define two different measures of the current flow: 
\[ J(\infty) \equiv \lim_{n \rightarrow \infty} J (nT) ~\mbox{and}~ J_{\mbox{sat}} \equiv \lim_{\omega\rightarrow\infty} \lim_{nT \rightarrow \infty} J(n T),\]
the first one will be called as the {\it asymptotic current} and the later one as the {\it saturated current}. The definition of the saturated current $J_{\mbox{sat}}$ demands a little bit of explanation. The limit $\omega\rightarrow\infty$ implies that $T\rightarrow 0$. So we have chosen $n \rightarrow\infty$ limit in such a manner that the time $nT$ also approaches towards $\infty$. Now onwards, all these measures of current will be considered only at the thermodynamic limit $L \rightarrow \infty$, where $L$ is the total number of lattice sites in the HCB chain. We analytically estimate the saturated current flow as
\begin{equation}
J_{\mbox{sat}} = \frac{2}{\pi}\,l_z^2\,\sin \nu = l_z^2\, J_{\mbox{in}},
\label{eq:Jsat-nu}
\end{equation}
where $J_{\mbox{in}}$ is the initial current given in Eq. \eqref{eq:Jin}.
\begin{figure}[b]
\includegraphics[height=7cm,width=8cm]{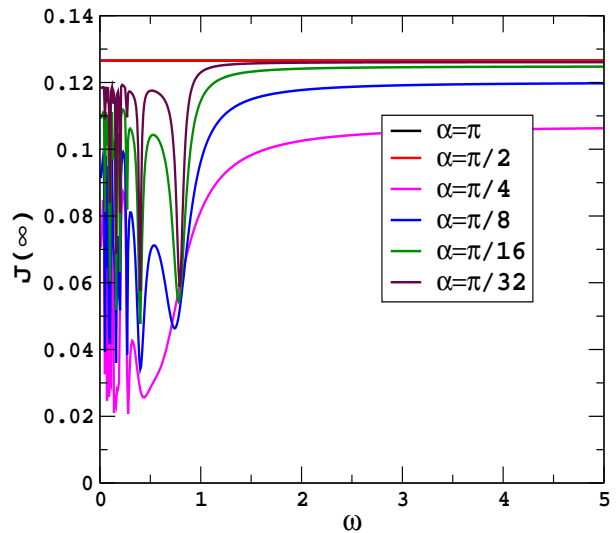}
\caption{(Color online) The asymptotic current versus the driving frequency $\omega$ is plotted for different values of the driving strength $\alpha$. Here we set $\Delta=0.3$ and the initial current is set by applying a twist $\nu = 0.2$. Here we see that the current is saturated at different finite values at large $\omega$ limit, but we do not see any DL.}
\label{fig:fig1}
\end{figure}
A recent study on the single kicked HCB chain has shown that, at large $\omega$ limit, $J_{\mbox{sat}} \sim \omega^{-3}$ and therefore the saturated current $J_{\mbox{sat}}$ disappears completely at $\omega \rightarrow \infty$ limit \cite{Tanay}. Moreover, this complete disappearance of the current was attributed to the DL. However, here, Eq. \eqref{eq:Jsat-nu} suggests that, for the double kick case, the saturated current $J_{\mbox{sat}}$ is independent of the driving frequency $\omega$. Therefore, we do not see a complete disappearance of $J_{\mbox{sat}}$, rather it saturates at different nonzero values depending on the parameter $\alpha, \Delta$, and $\nu$. This indicates the absence of any DL, and that leads to the survival of the saturated current under the driven staggered potential. Another important observation is that, unlike some previous studies including the case of the single kick HCB chain \cite{Klich,Tanay} where the saturated current was shown to follow $\nu^3$ law for $\nu \ll 1$, here Eq. \eqref{eq:Jsat-nu} suggests $J_{\mbox{sat}} \propto \nu$ for very small values of the twist $\nu$. 

We start our investigation by studying the asymptotic current $J(\infty)$ as a function of the driving frequency $\omega$ for different values of the kicking strength $\alpha$, and for a fixed value of the parameter $\Delta = 0.3$. This result is presented in Fig. \ref{fig:fig1}. In this figure, we note that for two special non-trivial values of the kicking strength $\alpha=\pi/2$ and $\pi$ (trivial case is $\alpha = 0.0$, i.e., no kick on the system), the initial current introduced by a twist $\nu = 0.2$ remains unchanged even at the asymptotic limit of the time for all values of the driving frequency $\omega$. We have mentioned earlier that, for $\alpha = \pi/2$, the effective Hamiltonian becomes $H_{\mbox{eff}} = (1-2\Delta) H_k = - 2 (1-2\Delta)\,\sigma_z \cos k$. Here, for any value of $\Delta$, the effective Hamiltonian is proportional to the free Hamiltonian $H_k$. On the other hand, for $\alpha = \pi$, the Floquet time-evolution operator becomes $e^{i 2T\sigma_z \cos k} = e^{-i H_k T}$, and therefore the corresponding effective Hamiltonian is exactly equal to the free Hamiltonian, i.e., $H_{\mbox{eff}} = H_k = - 2\sigma_z\,\cos k$. These imply that, for $\alpha = \pi/2$ and $\pi$, the effective Hamiltonians are proportional to the $z$-component of the pseudo-spin operator $\sigma_z$. Since the initial state $|\Psi(0)\rangle$ is a ground state of the twisted Hamiltonian $H_k^{(\nu)}$ which is also proportional to the operator $\sigma_z$, the initial state $|\Psi(0)\rangle$ is a stationary state for the time-evolution operators corresponding to the cases $\alpha = \pi/2$ and $\pi$. Consequently, the current flow in the system remains constant at its initial value under the time-evolution. For the other values of $\alpha$, dips and peaks are found in the asymptotic current for smaller values of the frequency $\omega < 1$, and depending on the kicking strength $\alpha$ the current flow reaches to a finite saturation value at the large $\omega$ limit. Later, we shall also show that, at the same limit of large $\omega$, the work-done on the system or the residual energy of the system saturates at a finite value. This implies that the HCB system has reached a {\it periodic steady state} where it stops absorbing any energy from the periodic driving \cite{Tanay,russomanno}. The saturated current shows its minimum value for $\alpha=\pi/4$. In comparison to the single kicked case, the asymptotic current for the double kicked case never reaches {\it zero} which indicates an absence of the DL, and that leads to the survival of current at the asymptotic limit of time. 

We are now going to study different properties of the saturated current $J_{\mbox{sat}}$ which is defined simultaneously at the large time limit ($nT \rightarrow \infty$) and at large frequency limit ($\omega \rightarrow \infty$). However, Fig. \ref{fig:fig1} clearly shows that the asymptotic current flow saturates around $\omega = 5$. The work-done on the system also saturates at the same value of $\omega$. Therefore, for all practical purposes, $\omega > 5$ can be considered as sufficiently large where the system reaches the periodic steady state. We set $\omega = 10$ for all the exact numerical calculations of the saturated current and work-done.

\begin{figure}[t]
\includegraphics[height=7cm,width=8cm]{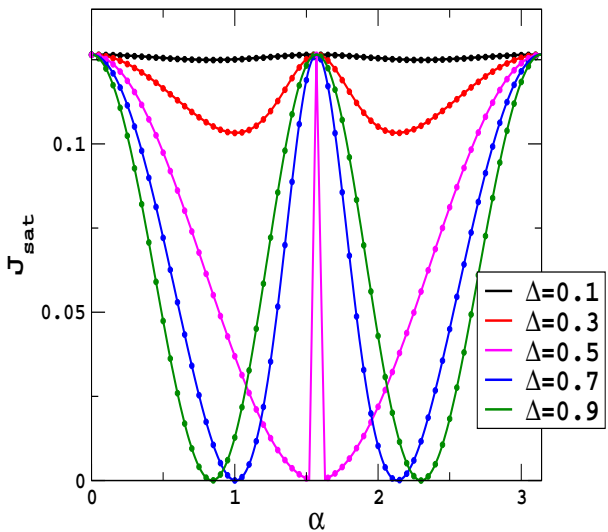}
\caption{(Color online) The saturated current $J_{\mbox{sat}}$ is plotted as a function of $\alpha$ for different values of the parameter $\Delta$. We set the initial twist $\nu=0.2$. Solid lines are the exact numerical results and the solid circles are representing our analytical estimation.}
\label{fig:fig2}
\end{figure}

In Fig. \ref{fig:fig2}, the saturated current $J_{\mbox{sat}}$ is plotted as a function of the driving strength $\alpha$ for different values of the parameter $\Delta$. Here we again set the twist $\nu = 0.2$. This figure shows a remarkable agreement between the exact numerical calculations (solid lines) and the analytical estimations (solid circles). For all values of $\Delta$, the saturated current varies periodically as a function of $\alpha$; and following our prediction, each of the curve shows mirror symmetry about $\alpha = \pi/2$, i.e., $J_{\mbox{sat}}$ is equal at $\alpha$ as well as at $\pi -\alpha$. From this figure, one can also observe that the saturated current remains fixed at its initial value for $\alpha = 0,\,\pi/2$ and $\pi$. For any values of $\Delta < 0.5$, we do not see any disappearance of the current flow for any value of the parameter $\alpha$. This is because, according to Eq. \eqref{eq:Jsat-nu}, the condition for observing $J_{\mbox{sat}} = 0$ is $l_z = 0$ and Eq. \eqref{eq:lz} suggests that we can not satisfy this condition for any real value of $\alpha$. Equation \eqref{eq:lz} also suggests that, for $\Delta = 0.5$, we have $l_z = \cos\alpha$ and therefore we get $J_{\mbox{sat}} = \cos^2\alpha\, J_{\mbox{in}}$. This suggests that, as we increase the kicking strength $\alpha$, the saturated current $J_{\mbox{sat}}$ decreases and becomes almost negligible when $\alpha \lesssim \pi/2$. But exactly at $\alpha = \pi/2$, as we discussed earlier, there is a discontinuity in the $J_{\mbox{sat}}$ versus $\alpha$ curve. For this case, we find $l_z = 1.0$ and thus $J_{\mbox{sat}} = J_{\mbox{in}}$. This discontinuity in $J_{\mbox{sat}}$ at $\alpha = \pi/2$ is reflected in the figure as a sudden jump from an almost zero current around $\alpha \lesssim \pi/2$ to the initial current $J_{\mbox{in}}$ and then an immediate drop from there to a very small current around $\alpha \gtrsim \pi/2$. Thereafter, the saturated current $J_{\mbox{sat}}$ starts increasing from that very small value and reaches once again to its initial value $J_{\mbox{sat}} = J_{\mbox{in}}$ at $\alpha = \pi$. For any given value of $\Delta$ within $0.5 < \Delta \leq 1.0$, we observe a complete disappearance of the saturated current at some values of $\alpha$ in the region $\alpha < \pi/2$ and also at $\pi - \alpha$ due to the mirror symmetry discussed earlier. These pair of values of $\alpha$ are determined by setting $l_z = 0$ and from Eq. \eqref{eq:lz} we get the condition:
\begin{equation}
\begin{split}
\alpha & = \pm \cot^{-1}\left(\sqrt{2\Delta -1}\right)\\ & = \cot^{-1}\left(\sqrt{2\Delta -1}\right)\,\,\mbox{and}\,\,\pi - \cot^{-1}\left(\sqrt{2\Delta -1}\right).
\end{split}
\end{equation}
For $\Delta = 0.7$, we get from the above condition that the saturated current disappears at $\alpha = \pm\cot^{-1} (\sqrt{0.4}) \simeq 1.007\,\,\mbox{and}\,\, 2.135$. Similarly, for $\Delta = 0.9$, we observe the disappearance of $J_{\mbox{sat}}$ at $\alpha \simeq 0.841 \,\,\mbox{and}\,\, 2.301$. Fig. \ref{fig:fig2} confirms this analytical estimation.

\begin{figure}[b]
\includegraphics[height=7cm,width=8cm]{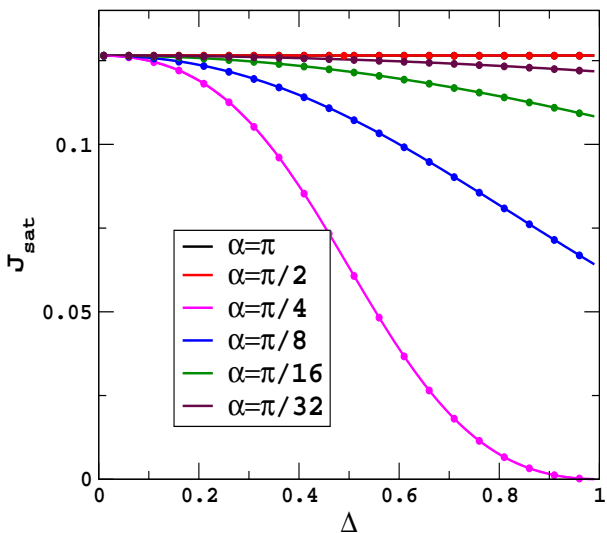}
\caption{(Color online) The saturated current $J_{\mbox{sat}}$ versus the time interval between two kicks determined by the parameter $\Delta$ is plotted for different values of the kicking strength $\alpha$. Solid lines are representing exact numerical calculation whereas the circles are representing our analytical estimation.}
\label{fig:fig3}
\end{figure}

\begin{figure}[t]
\includegraphics[height=7cm,width=8cm]{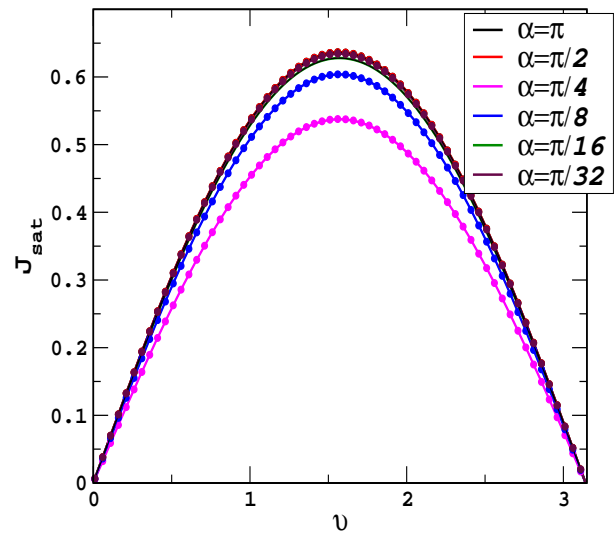}
\caption{(Color online) The saturated current $J_{\mbox{sat}}$ is plotted as a function of the initial twist $\nu$. As we discussed in the text, here $J_{\mbox{sat}} \propto \sin\nu$. This suggests that, unlike the observed ``$\nu^3$-law" in case of the singled kicked HCB system, here the double-kicked HCB system follows $J_{\mbox{sat}} \propto \nu$ at very small twist $\nu \ll 1$.}
\label{fig:fig4}
\end{figure}

We now study how the saturated current can be tuned by varying the time interval between the two kicks determined by the parameter $\Delta$ for different fixed values of the kicking strength $\alpha$. In Fig. \ref{fig:fig3}, we have presented this result. This figure again shows a very good agreement between the exact numerics and our analytical estimation. Here, the very first thing which we observe is that, at $\Delta = 0.0$, the saturated current is same for all values of the driving strength $\alpha$. This we can easily explain from the diagram in Fig. \ref{fig:fig0}. From this diagram, we see that at $\Delta = 0.0$ both the kicks are acting on the system at the same time. Since, the driving strength of the kicks are of opposite sign, then they naturally cancel each other. That means, the dynamics of the system is governed by the free Hamiltonian and hence the effective Hamiltonian $H_{\mbox{eff}} = H_k = - 2 \sigma_z \cos k$. This we can also obtain mathematically by substituting $\Delta = 0.0$ in the Floquet time-evolution operator given in Eq. \eqref{eq:FkT}. Therefore, due to this effective absence of the $\delta$-function kicks, the initial current carrying ground state $|\Psi(0)\rangle$ remains stationary under the time-evolution and consequently the current flow in the system remains fixed at the initial value $J_{\mbox{in}} = (2/\pi)\sin\nu \simeq 0.126$ irrespective of the driving strength $\alpha$. The next we see that, for $\alpha=\pi/2$ and $\pi$, the saturated current is frozen at its initial value $J_{\mbox{in}} \simeq 0.126$ for all values of $\Delta$. Therefore, for these cases, we can not alter $J_{\mbox{sat}}$ by tuning $\Delta$. For the other values of $\alpha$, the saturated current decreases as $\Delta$ increases. Since the saturated current is remained fix at its initial value for $\alpha = \pi/2$ and $\pi$; we expect that for $\alpha=\pi/4$, the intermediate value between $\pi/2$ and $\pi$, the saturated current should show maximum drop as a function of $\Delta$. Fig. \ref{fig:fig4} is showing that the saturated current indeed decreases at the fastest rate when $\alpha=\pi/4$ and it approaches to zero as the parameter $\Delta$ approaches to $1.0$. At $\Delta = 1.0$, the dynamics is very much different which we can see from the expression of the time-evolution operator $\mathcal{F}_k(T)$ given in Eq. \eqref{eq:FkT}. According to this equation, at $\Delta = 1.0$, we have
\begin{equation}
\mathcal{F}_k(T) = \exp\left(i\, \alpha \sigma_x\right) \exp\left(i\, 2\, T \sigma_z \cos k\right) \exp\left(- i\, \alpha \sigma_x \right).
\end{equation} 
Therefore, for a time-evolution $t = nT$, an integer multiple of the time-period $T$, the time-evolution operator becomes
\begin{equation}
\begin{split}
\mathcal{F}_k(nT) &= \mathcal{F}_k(T)^n = \exp\left(i\, \alpha \sigma_x\right) \exp\left(i\, 2\, n T \sigma_z \cos k\right)\\ &\times \exp\left(- i\, \alpha \sigma_x \right).
\end{split}
\end{equation}
The above expression suggests that, in addition to an initial and a final kick, the dynamics of the system is governed by the free Hamiltonian $H_k = -2\sigma_z\,\cos k$ for the whole time-evolution $t=nT$. This dynamics is fundamentally different from the double-kicked system. Hence, here we ignore this particular case and all the results are presented for $\Delta < 1.0$. For other values of $\alpha$, the saturated current does not reach so close to zero for $\Delta \lesssim 1.0$. 
 
Finally, the saturated current $J_{\mbox{sat}}$ is investigated as a function of the initial current. As we mentioned several times, the initial current flow in the system is instigated by applying a twist $\nu$ in the system. Therefore, in Fig. \ref{fig:fig3}, instead of plotting the saturated current as a function of the initial current, we have presented exact numerics together with the analytical estimation for the saturated current as a function of the twist $\nu$ for different values of the kicking strength $\alpha$ where $\Delta$ is fixed at $0.3$. This initial current saturates to some nonzero value under the periodically driven double kicked staggered onsite potential. This happens for any value of the kicking strength $\alpha$. From Fig. \ref{fig:fig3}, we can conclude that, as we increase the initial current in the system, the saturated current also increases, and it reaches its maximum value at $\nu=\pi/2$. As we further increase the twist, then the initial current decreases and we see a drop in the saturated current $J_{\mbox{sat}}$. The saturated current reaches its minimum value {\it zero} when $\nu = \pi$. Importantly, for the fixed values of $\alpha$ and $\Delta$, $J_{\mbox{sat}} \propto J_{\mbox{in}}$ and therefore $J_{\mbox{sat}} \propto \sin\nu$. This suggests an important result that, for a very small twist $\nu \ll 1$, the saturated current $J_{\mbox{sat}} \propto \nu$. This behavior is fundamentally different from the single kicked HCB model where $J_{\mbox{sat}} \propto \nu^3$ was reported when $\nu \ll 1$ \cite{Tanay}.  

\begin{figure}[b]
\includegraphics[height=7cm,width=8cm]{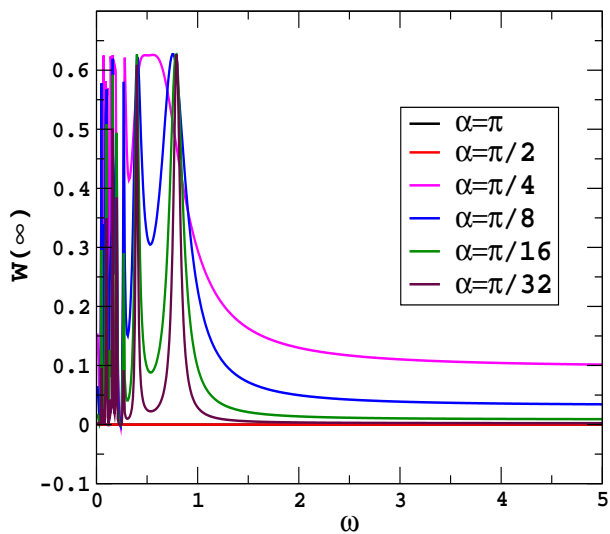}
\caption{(Color online) The asymptotic work-done $W(\infty)$ is plotted as a function of $\omega$ for different values of the driving strength $\alpha$. Here we set the values of the other parameters same as given in Fig. \ref{fig:fig1}.}
\label{fig:fig6}
\end{figure}

\subsection{Work-done}

The work-done on the system $W_d$ (also known as the residual energy \cite{maity}) is calculated using Eq. \eqref{eq:wd_defn2}. From this expression we find that 
\begin{equation}
W_d(nT) = \frac{2}{L} \sum_k \bigl[1-f(k)\bigr] \langle\Psi_{k}(0)|\sigma_{z}|\Psi_{k}(0)\rangle \cos k,
\end{equation}
where the function $f(k)$ is given in Eq. \eqref{eq:fk}. Here again, at the thermodynamic limit $L \rightarrow \infty$, we can write 
\begin{equation}
W_d(nT) = \frac{1}{\pi}\int\limits_{-\pi/2}^{\pi/2} dk\, \bigl[1-f(k)\bigr] \langle\Psi_{k}(0)|\sigma_{z}|\Psi_{k}(0)\rangle \cos k.
\end{equation}
Following the two definitions of the current at the asymptotic limit $n \rightarrow \infty$, here we also define two different measures of the work-done:
\[W(\infty) \equiv \lim_{n\rightarrow \infty} W_d(nT) ~\mbox{and}~ W_{\mbox{sat}} \equiv \lim_ {\omega\rightarrow\infty}\lim_{nT\rightarrow\infty} W_d(nT).\]
Here we also follow the same nomenclature: $W(\infty)$ as the asymptotic work-done on the system and $W_{\mbox{sat}}$ as the saturated work-done. At the thermodynamic limit $L\rightarrow \infty$, our analytical estimation gives a simple relation for the saturated work-done $W_{\mbox{sat}}$ as:
\begin{equation}
W_{\mbox{sat}} = \frac{2}{\pi} \left(1-l_z^2\right)\,\cos \nu = \frac{2}{\pi}\, l_y^2\,\cos \nu,
\end{equation}
where the parameters $l_y$ and $l_z$ are given in Eq. \eqref{eq:lz}.

\begin{figure}[t]
\includegraphics[height=7cm,width=8cm]{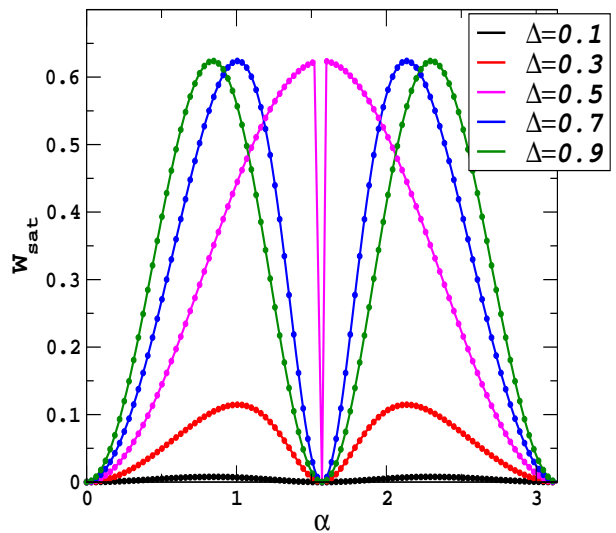}
\caption{(Color online) The saturated work-done $W_{\mbox{sat}}$ versus the driving strength $\alpha$ is plotted for different values of $\Delta$. The initial twist $\nu = 0.2$. Solid lines and the circles are respectively representing the exact numerical calculation and the analytical estimation.}
\label{fig:fig7}
\end{figure}

\begin{figure}[b]
\includegraphics[height=7cm,width=8cm]{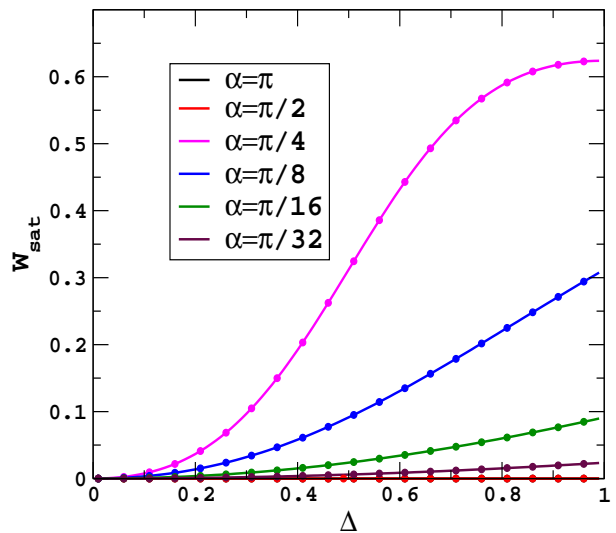}
\caption{(Color online) The saturated work-done $W_{\mbox{sat}}$ is plotted as a function of the parameter $\Delta$ which determines the time interval  between two kicks for different values of the driving strength $\alpha$. The initial twist $\nu = 0.2$. Solid lines and the circles are respectively representing the exact numerical calculation and the analytical estimation.}
\label{fig:fig8}
\end{figure}

In Fig. \ref{fig:fig6}, the asymptotic work-done on the system $W(\infty)$ is plotted as a function of the frequency $\omega$ for different values of the driving strength $\alpha$. Here we set the parameter which determines the time interval between the two kicks at $\Delta = 0.3$ and the initial twist at $\nu = 0.2$. For $\alpha=\pi$ and $\pi/2$, the work-done on the system is {\it zero}. This result agrees very well with our expectations. Because of these cases, the HCB chain remains in the superfluid phase, and thus there is no work to be done on the system to maintain the current flow. For the other values of the driving strength $\alpha$, the periodically kicked staggered potential introduces an energy gap (system becomes Mott insulator) in the system at the time of each kicking, and that leads to a drop in the current. Therefore, we need to do some work on the system to steady the current flow. Similar to the behavior of the asymptotic current $J(\infty)$, the work-done at the asymptotic limit $W_d(\infty)$ also shows peaks and dips up to $\omega = 1$. However, we observe an anti-correlation between the peaks and dips of $J(\infty)$ and $W_d(\infty)$: wherever $J(\infty)$ shows a peak, then $W_d(\infty)$ shows a dip exactly at the same value of $\omega$ and vice versa.  For $\omega > 1$, the work-done starts to saturate, and it reaches its saturation value around $\omega \gtrsim 5$. This behavior of $W_d(\infty)$ is consistent with the behavior of the current flow as a function of the driving frequency $\omega$. As we discussed earlier, this saturation implies that the system has reached a periodic steady-state \cite{russomanno}. As a consequence, the system stops absorbing any energy from the driving.

We now study the saturated work-done $W_{\mbox{sat}}$ as a function of the kicking strength $\alpha$ for different time intervals between the two kicks determined by the parameter $\Delta$. Here we fix the initial current by setting the twist at $\nu = 0.2$. We present this result in Fig. \ref{fig:fig7}. If we compare this result with the result of $J_{\mbox{sat}}$ versus $\alpha$ presented in Fig. \ref{fig:fig2}, we observe that the lower value of $J_{\mbox{sat}}$ demands higher value of $W_{\mbox{sat}}$ and vice versa. We explain this observation by the following argument: as we increase the driving strength $\alpha$ from {\it zero} to $\pi/2$, the energy gap created at the time of each kick also increases and that leads to a smaller current flow in the system. Therefore, to maintain a steady current flow, one has to do more work on the system. The same reasoning not only explains the behavior of the current flow in the system and the work-done on the system in the parameter regime $\pi/2 < \alpha < \pi$, but it also explains consistently the following observations on the saturated work-done on the system $W_{\mbox{sat}}$: (i) it remains fixed at {\it zero} when the saturated current $J_{\mbox{sat}}$ remains constant for all values of the driving strength $\alpha$; and (ii) it reaches its maximum value when the saturated current almost vanishes.

In Fig. \ref{fig:fig8}, we present the behavior of $W_{\mbox{sat}}$ as a function of the parameter $\Delta$ for different values of $\alpha$ where the initial twist is fixed at $\nu = 0.2$. Following our expectation, here also we see qualitatively similar properties of $W_{\mbox{sat}}$. For example: (i) the saturated work-done on the system $W_{\mbox{sat}} = 0$, whenever the current $J_{\mbox{sat}} = J_{\mbox{in}}$; (ii) The saturated work-done becomes non-zero whenever the saturated current drops from the initial current, whether that due to the variation of the driving strength $\alpha$ or the variation of the parameter $\Delta$. Therefore, a general observation is that a larger drop in the saturated current from the initial current leads to a larger saturated work-done on the system and vice versa.

\subsection{Time-evolution of density of particles}

Figure \ref{fig:fig9} shows the dynamics of the density of particles as a function of the stroboscopic time $t = nT$ (along $x$-axis) and the site location $l$ (along $y$-axis). Here we fix the driving strength $\alpha = \pi/2$. We have chosen this particular value of $\alpha$ because, at this value of the driving strength, the DL was observed in case of the single kicked HCB model \cite{Tanay}. Following this recent work, here we have chosen a $200$ lattice sites HCB system. At the initial time $t=0.0$ or $n=0$, we have placed one particle each at sites $51$ to $150$ and kept the remaining sites empty. As the time $t=nT$ evolves, the particles are expected to spread out with the maximum of the absolute value of group velocities $v_k^\pm$. The group velocities can be obtained from the quasi-energies $\mu_k^\pm$ using the relation $v_k^\pm = d\mu_k^\pm/dk$. At $\alpha = \pi/2$, we obtain from Eq. \eqref{eq:mu_k} that $\mu_k^\pm = \pm\, 2 (1-2\Delta) \cos k$, and consequently we calculate $v_k^\pm = \mp\, 2 (1-2\Delta) \sin k$. The maximum of the absolute value of group velocity will be obtained by substituting $\sin k = 1.0$, and thus we get max$\left(|v_k^\pm|\right) = 2 (1-2\Delta)$ when $\Delta \leq 0.5$. Here we are not going to study the cases for which $\Delta > 0.5$ because the quasi-energy $\mu_k^\pm$ has a symmetry about $\Delta = 0.5$, i.e., if we substitute $\Delta$ by $1-\Delta$ in Eq. \eqref{eq:mu_k}, we get the same quasi-energies.  

As we observe the spreading of the particles stroboscopically, we find that the boundaries separating the occupied and unoccupied regions are spreading linearly in light-cone like fashion with a rate determined by max$\left(|v_k^\pm|\right)$. Therefore, the dynamical equation for this linear spreading in the configuration space or lattice space will be:
\begin{equation}
\frac{dl}{dt} = \frac{1}{T} \frac{dl}{dn} = \pm\, \max\left(\left|v_k^\pm\right|\right) = \pm\, 2 (1-2\Delta).
\end{equation}
For a particle initially placed at some lattice point $l_0$, here $51 \leq l_0 \leq 150$, then at any arbitrary instant $n$, the particle will be at the lattice site
\begin{equation}
l = l_0 \pm 2(1-2\Delta)\, nT.
\end{equation}
In Figs. \ref{fig:fig9}(a)-(f), we have shown these linear spreading of the boundaries by solid red lines and these are obtained by plotting the trajectories of the two particles placed initially at the boundary sites $51$ and $150$. For larger values of $T\, (= 1.0, 10.0)$, we see that the particles have reached the boundaries of the HCB lattice $(l = 200)$ very quickly; and then they reflected back from there and form an interference patterns. These we see for both $\Delta = 0.1$ and $0.3$. Therefore, these interference patterns are appearing purely due to the effect of finite size lattice $(L = 200)$ and we will not see such patterns if we go to the thermodynamic limit $L \rightarrow \infty$. 

In Figs. \ref{fig:fig9}(g)-(i), we set the parameter $\Delta = 0.5$. For this case, the particles do not move from their initial position with time-evolution, i.e., $l = l_0$ for all $n$, and this is valid for all $T$. We have discussed earlier that, the effective Hamiltonian becomes a {\it null} operator when $\alpha = \pi/2$ together with $\Delta = 0.5$ and that leads to the absence of any dynamics in the system. Consequently, the particles placed in the region $51 \leq l \leq 150$ do not move from their initial position and this we observe in Figs. \ref{fig:fig9}(g)-(i). Even though these figures look similar to the situation when the DL prevents the particle from spreading \cite{Tanay}, but here the origin of the localization is due to a very trivial reason, and that is the absence of any dynamics. Here we are not presenting any results corresponding to the other values of the driving strength $\alpha$, because for all those cases the dynamics of the density of particles qualitatively be the same, i.e., for any values of $\Delta$, we will not observe the absence of dynamical evolution (i.e., $v_k^\pm \neq 0.0$), and that leads to the light-cone like spreading of the density profile for all cases.  

\begin{widetext} 
\begin{center}
\begin{figure}[t]
\includegraphics[height=10cm,width=18cm]{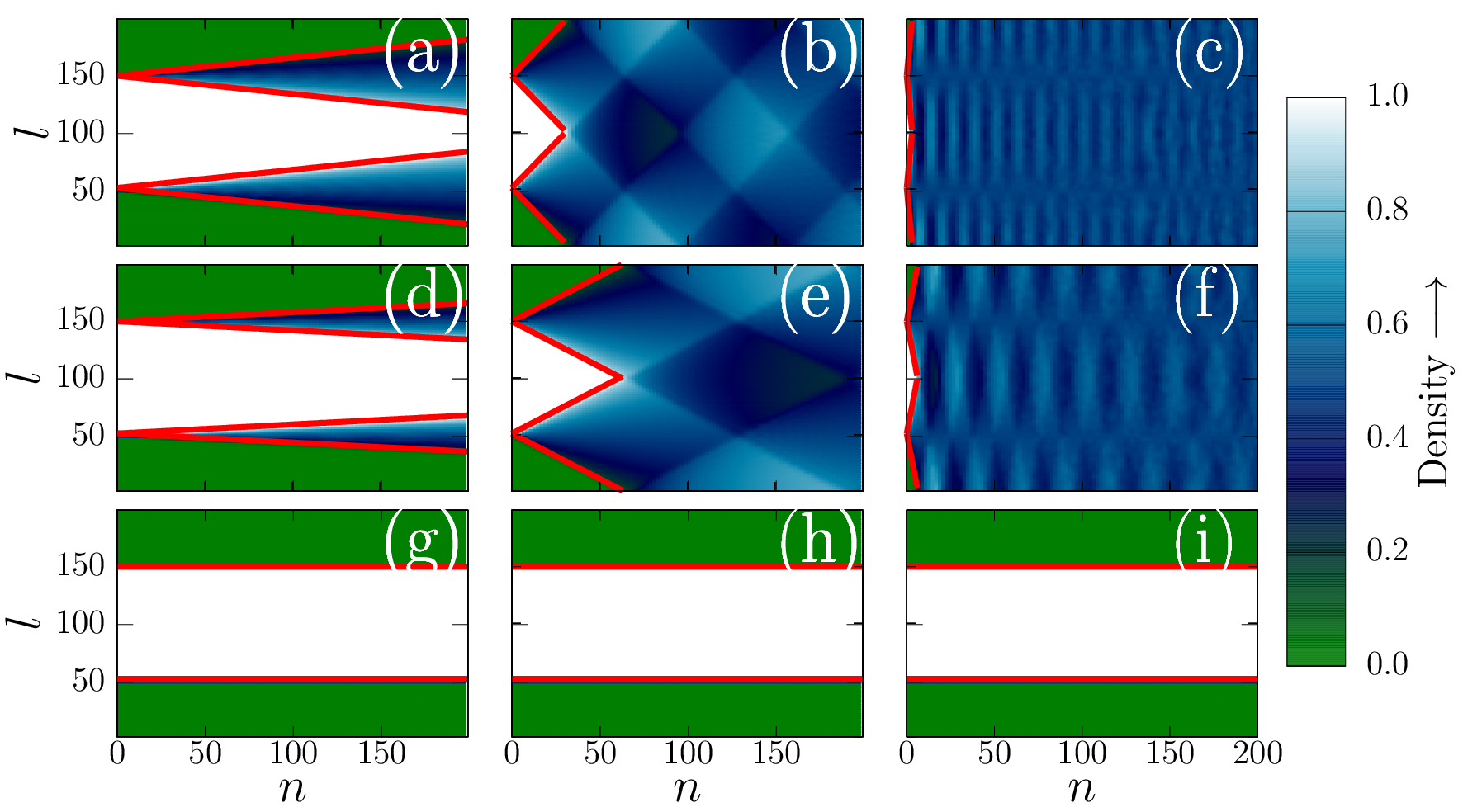}
\caption{(Color online) The time-evolutions of the density of particles placed in a HCB system of $200$ lattice sites are shown as a function of the stroboscopic time $t = nT$ for a fixed value $\alpha = \pi/2$. Here, as we go from the left to the right, the time-period $T$ varies; and when we go from the top to the bottom, the parameter $\Delta$ varies: (a) $\Delta = 0.1,\, T = 0.1$; (b) $\Delta = 0.1,\, T = 1.0$; (c) $\Delta = 0.1,\, T = 10.0$; (d) $\Delta = 0.3,\, T = 0.1$; (e) $\Delta = 0.3,\, T = 1.0$; (f) $\Delta = 0.3,\, T = 10.0$; (g) $\Delta = 0.5,\, T = 0.1$; (h) $\Delta = 0.5,\, T = 1.0$; and (i) $\Delta = 0.5,\, T = 10.0$. Solid red lines are showing the boundaries of the light-cone like regions.}
\label{fig:fig9}
\end{figure}
\end{center}
\end{widetext} 

\section{Conclusion}
\label{conclude}

In this paper, we study the dynamics of a current-carrying initial state in an HCB chain experiencing two $\delta$-function kicks within a single time-period $T$ in the form of a staggered onsite potential. Unlike the singled kicked HCB chain, here we do not see any onset of dynamical localization at the long-time limit $n \rightarrow \infty$. Rather we have observed that, depending on the time interval between the two kicks determined by the parameter $\Delta$ and the driving or kick strength $\alpha$, the initial current saturates at some finite nonzero values. For three special values of $\alpha$, such as $0,\, \pi/2$ and $\pi$, the initial current does not change with the time-evolution. Our study of the work-done on the system as a function of the frequency $\omega$ reveals that whenever the current gets saturate, then the work-done also stops varying and reaches its saturation value. This indicates that the system has reached a periodic steady state and it has stopped absorbing any energy from the periodic driving.

Further studies have shown a general trend that for larger saturated current the work-done saturates at a smaller value and vice versa. For the cases mentioned above, when the initial current does not change with time-evolution, we see zero work-done throughout the whole time-evolution. This is a generic property that was also observed in the singled kicked case. Finally, we have also observed a light-cone like spreading of the density of particles in the real space when we have placed the hard-core localized bosons in one part of the chain. Here we have particularly placed the bosons at the central part of the lattice. The maximum value of the group velocity determines the rate of this light cone like spreading of the particles.   

\acknowledgments

J.N.B thanks DST-SERB, India, for support through Project No. EMR/2016/003289.

\end{document}